\documentclass{pasa_mod}
\usepackage[numbers]{natbib}

\usepackage{aas_macros}
\usepackage{hyperref} 
\hypersetup{colorlinks,citecolor=blue,linkcolor=blue,urlcolor=blue}
\hypersetup{draft}

\usepackage{newtxtext,newtxmath}

\usepackage{graphicx}	
\usepackage{amsmath}	
\usepackage{amssymb}	
\usepackage[export]{adjustbox}
\usepackage{subfig}

\title[An all-photonic focal-plane wavefront sensor]{An all-photonic focal-plane wavefront sensor}

\author[B. R. M. Norris et al.]{
Barnaby R. M. Norris,$^{1,2,3}$\thanks{E-mail: barnaby.norris@sydney.edu.au}
Jin Wei,$^{1,2,4}$
Christopher H. Betters$^{1,2,4}$,
Alison Wong$^{1,2}$
and Sergio G. Leon-Saval$^{1,2,4}$

\affil{$^{1}$Sydney Institute for Astronomy, School of Physics, Physics Road, University of Sydney, NSW 2006, Australia}
\affil{$^{2}$Sydney Astrophotonic Instrumentation Laboratories, Physics Road, University of Sydney, NSW 2006, Australia}
\affil{$^{3}$AAO-USyd, School of Physics, University of Sydney 2006}
\affil{$^{4}$Institute of Photonics and Optical Science, School of Physics, University of Sydney 2006}
}

\begin{document}
\label{firstpage}

\begin{frontmatter}
\maketitle

\begin{abstract}
Adaptive optics (AO) is critical in astronomy, optical communications and remote sensing to deal with the rapid blurring caused by the Earth's turbulent atmosphere. But current AO systems are limited by their wavefront sensors, which need to be in an optical plane non-common to the science image and are insensitive to certain wavefront-error modes. Here we present a wavefront sensor based on a photonic lantern fibre-mode-converter and deep learning, which can be placed at the same focal plane as the science image, and is optimal for single-mode fibre injection. By measuring the intensities of an array of single-mode outputs, both phase and amplitude information on the incident wavefront can be reconstructed. We demonstrate the concept with simulations and an experimental realisation wherein Zernike wavefront errors are recovered from focal-plane measurements to a precision of $5.1\times10^{-3}\;\pi$ radians root-mean-squared-error. 
\end{abstract}


\end{frontmatter}



\section*{Introduction}
\label{sec_intro}
Due to the blurring caused by the Earth's atmosphere as starlight passes through it, adaptive optics has become central to the advance of modern astronomy, including the imaging of extra-solar planets, newly-forming planetary systems, dying stars and active galactic nuclei. It also offers key advantages in fields where any type of distorted media hinders the detection and/or manipulation of the desired optical signal such as free-space optical communications, remote sensing, in-vivo imaging and manipulation of living cells.

Excellent reviews of adaptive optics systems are given in \cite{Davies2012} and \cite{Guyon2018}. In an adaptive optics system, a deformable mirror (DM) situated at the telescope pupil plane is used to rapidly apply corrections to the incident wavefront, cancelling out the effect of atmospheric turbulence. Modern DMs consist of thousands of electrically driven actuators, each applying a small deformation to the mirror surface on time scales of milliseconds. The performance of this method thus largely depends on how accurately the current state of the wavefront is known -- a task accomplished (in conjunction with various reconstruction algorithms) by the system's wavefront sensor (WFS).

While the goal of the AO system is to produce the optimal image in the instrument's focal-plane, the current state of the wavefront can not easily be determined from this focal-plane image alone. This is because the measured image (obtained by an imaging detector such as a CCD or CMOS chip) contains information only on the intensity of the beam, and is missing the phase information. But phase information is crucial in measuring the incident wavefront. For this reason, AO systems have conventionally used a separate wavefront sensor, positioned in a separate pupil plane (usually reimaged via a dichroic beamsplitter) rather than at the image plane. There exist several designs for these pupil-plane wavefront sensors, such as the Shack-Hartmann wavefront sensor \citep{Platt2001}, the pyramid wavefront sensor \citep{Ragazzoni1996} and the curvature wavefront sensor \citep{Roddier1988b}. 

Systems solely using pupil-plane wavefront sensors have some important disadvantages. Firstly, they are subject to non-common path aberrations -- differences between the wavefront seen by the WFS and that used to make the image, due to the non-common optical components traversed by the wavefront-sensing and science beams \citep{Sauvage2007}. Since these aberrations are not seen by the wavefront sensor, they are not corrected, and this is currently the main limiting factor in the performance of high-contrast Extreme-AO systems in astronomy \citep{NDiaye2018}. It can take the form of both low-order aberrations (particularly harmful when a coronagraph is used) and high-order ones, which can produce static and quasi-static speckle. The latter is particularly insidious since it slowly varies depending on telescope pointing and other parameters, so can not easily be calibrated for. 

Another major disadvantage is that there exist some highly detrimental aberrations to which pupil-plane WFSs are insensitive, specifically the so called Low Wind Effect (LWE) or Island Effect \citep{Sauvage2016, Milli2018, NDiaye2018, Vievard2019}. This arises due to phase discontinuities across the secondary-mirror support structure in the telescope pupil, exacerbated by thermal effects that these structures create when the wind is low. Since this takes the form of a sudden step in phase across a region obscured (by the mirror support structures) in the pupil plane, they are virtually invisible to a pupil-plane WFS. However they have an extremely strong effect in the image plane, and are also a limiting factor in the performance of adaptive optics systems. 

For these reasons, a focal-plane wavefront sensor (FP-WFS) has been long desired. As mentioned, a simple image will not do, since this does not contain any phase information. This missing information results in an ambiguity in any inferred wavefront determination. However various ingenious methods have been devised to address this, each with their own advantages and limitations. Phase diversity methods \citep{Gonsalves1982} generally rely on a set of two simultaneous images, taken with different aberrations (for example, both an in-focus and defocused image), allowing the ambiguity to be broken. However this requires some physical method to produce these two images, and also (due to the highly nonlinear relationships involved) relies on computationally expensive iterative algorithms that preclude real-time operation. An analytic solution (Linearized Analytic Phase Diversity) has been developed to allow real-time operation \citep{Mocoeur2009, Vievard2018} but this relies on a linear approximation, requiring the magnitude of phase aberrations be small (<<1 radian), a condition that aberrations such as the Low Wind Effect does not necessarily fulfil. Other methods, such as the Fast \& Furious method \citep{Korkiakoski2014} avoid the need for a simultaneous, aberrated image by using knowledge of the DM state, but also rely on a linear approximation. The Zernike Asymmetric Pupil Wavefront Sensor \citep{Martinache2013} is based on a kernel-phase analysis of the focal-plane image, and addresses the lack of phase information by inserting an asymmetric obstruction in the telescope pupil. It also relies on a linear approximation. Another class of methods rely on actively modulating the DM to generate `probe' speckles, which are then modulated in an iterative fashion to break phase ambiguity \citep{Martinache2014}.

Furthermore, all these focal-plane wavefront sensors have a major disadvantage -- they assume that an imaging detector of some sort, with sufficient readout speed, is present at the focal plane. However for advanced exoplanet applications a spectrum of the exoplanet is desired, to allow characterisation of the composition of exoplanet atmospheres, mapping via doppler shift from planet rotation and even the detection of biological signatures \citep{Snellen2015, Wang2018, Crossfield2014, Seager2010}. This requires that rather than using an imaging detector, the planet image be fed to a high-dispersion spectrograph, either via injection into an optical fibre located at the image plane or by conventional optical means.

In this paper, we present a type of focal-plane wavefront sensor that directly measures the phase as well as intensity of the image, without any linear approximations or active modulation. Leveraging photonic technologies as well as machine learning, the Photonic Lantern Wavefront Sensor (PL-WFS) uses a monolithic photonic mode converter known as a photonic lantern (PL) to determine the complex amplitude of the telescope point-spread function (PSF), via the conversion of multi-modal light into a set of single-mode outputs, as depicted in Figure \ref{fig_rsoftphasediag} (top). The desired wavefront information can be determined by simply measuring the intensity of each of the single-mode outputs, which are also ideal for injection into a high-dispersion, diffraction-limited spectrograph, ideal for exoplanet characterisation \citep{Betters2016}. 
In previous efforts a photonic lantern was simulated to measure the tip and tilt of an injected beam \citep{Corrigan2016, Corrigan2018}, but now higher order terms describing the shape of the wavefront can be actually measured.

Since the relationships between input phase and output intensities is non-linear, a deep neural network is used to perform the reconstruction. These deep learning methods \citep{Lecun2015} have recently exploded in popularity across many fields of science and engineering. In essence, a neural network learns the relationship between the inputs (in this case wavefront phase) and outputs (in this case the intensities of the single-mode core lantern outputs) of some system. Then, given a new, previously unseen set of outputs, it can infer what the input is. The use of simple neural networks for multimode fibre applications has been investigated for several decades, including for image categorisation \citep{Aisawa1991} and information transition \citep{Marusarz2001}. However recent advances in computational power and deep learning methods have allowed more complex applications to multimode fibres, such as convolutional neural networks, to be investigated \citep{Rahmani2018}. Another advantage of such methods is that they can perform the required inferences extremely quickly, with currently available frameworks able to perform highly complex, true non-linear inferences with sub-millisecond latency \citep{nvidia2020}.

\section*{Results}

\subsection*{Numerical simulation and theory}
\label{sec_theory}

\begin{figure*}
\centering
\includegraphics[width=0.9\textwidth,valign=c]{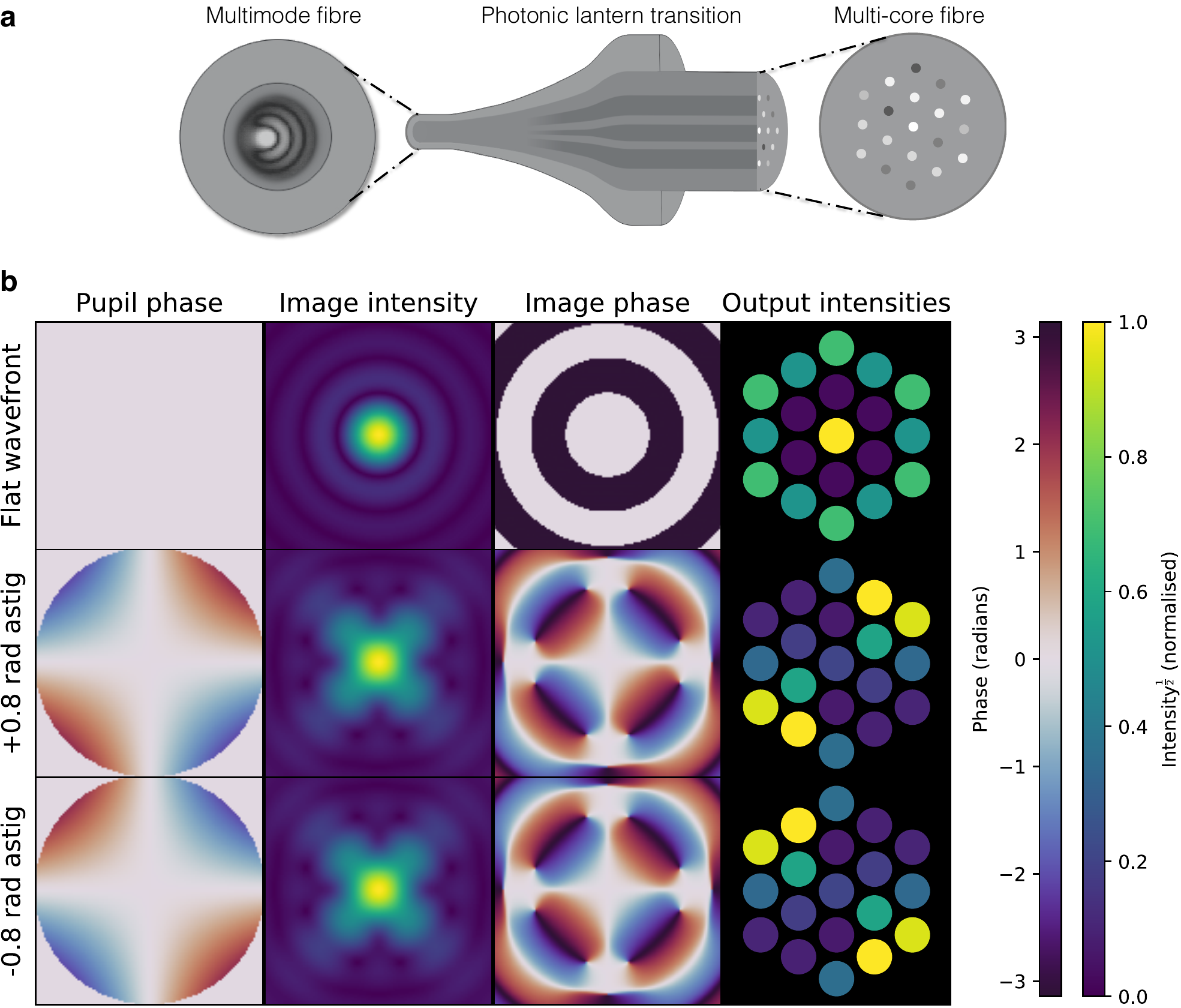}
    \caption{{\bf Non-degenerate response of the photonic lantern wavefront sensor to focal plane phase. 
    } {\bf a} Schematics of a multi-core photonic lantern showing how the phase and intensity of the input field into the multimode fibre end-face evolve into an array of uncoupled single-mode cores with different intensities. {\bf b} The results of three {\ttfamily RSoft} simulations demonstrating the concept of the photonic lantern wavefront sensor, and its ability to measure both amplitude and phase. The first column shows the phase of the wavefront, and the second and third columns show the intensity and phase of the resulting PSF respectively. The fourth column shows the intensities of the 19 single-mode outputs of the photonic lantern, when the corresponding PSF is injected. In the first example (first row) a flat wavefront is used. In the second and third rows, astigmatism with an amplitude of 0.8 radians, but with opposite signs, is introduced. This results in identical intensity structure in the image plane (2nd column), and so could not be distinguished with an imaging sensor. However the (usually un-measured) phase in the focal plane (3rd column) shows the difference between the two astigmatism terms, which is successfully measured by the photonic lantern (as shown by the different set of outputs from the lantern, in the 4th column). Simulations are performed at a wavelength of 1550~nm. Intensities are plotted with a square-root stretch to better show faint detail.}
    \label{fig_rsoftphasediag}
\end{figure*}

Since the modes excited within a multimode fibre (MMF) are a function of the electric field at the input, by measuring the relative power in each mode at the fibre's output it is in principle possible to reconstruct spatial information describing the input beam. Although power mixes between the various modes of the fibre as it propagates, as long as the fibre remains unperturbed (e.g. by strain or temperature) then the relationship between the input and output mode fields can be determined. This principle has allowed the development of basic imaging applications, wherein an image projected into the input face of the fibre is reconstructed by imaging the output mode field \citep{Cizmar2012}. Although a simple intensity image of the PSF does not contain the necessary information to reconstruct the wavefront, the combination of modes excited within a MMF is a function of both the phase and the amplitude of the incoming light. Hence if the power in each mode of the fibre is known, it should be possible to infer the complex wavefront of an injected PSF. 

In standard astronomical fibre-based spectroscopy, the point-spread function (PSF) of the telescope while observing a star is indeed injected into a multimode fibre. However reconstructing the complex wavefront by simply imaging the output of the MMF is difficult for a number of reasons. Firstly, the relationship between the modes at the input and output (the transfer function) is not constant, since the fibre, existing in the relatively hostile environment of a working observatory, will be subjected to various changes in strain and temperature. Secondly, in astronomical applications the light levels involved are extremely low, and so imaging the output mode field onto the many (read-noise limited) pixels of a CCD or CMOS detector -- operated at 1000s of frames/second -- is problematic. Thirdly, the decomposition of a mode field image into a set of coefficients of each mode is a complicated, computationally expensive and delicate task, not suited to the high degree of robustness and low latency required in a working observatory. Finally, if the output light from the fiber is allowed to propagate in free space to a camera it is difficult to effectively use the same light (at another wavelength) for science measurements, such as in a high-resolution spectrograph.

These issues can be addressed by taking advantage of a photonic mode converter known as a photonic lantern (PL) \citep{Leon-Saval2005, Leon-Saval2013, Birks2015}. A photonic lantern acts as an interface between a multimode fibre and multiple single-mode fibres. By way of an adiabatic taper transition, light contained within the multimode fibre is efficiently transferred into a discrete array of single-mode outputs as seen in Figure \ref{fig_rsoftphasediag} (top). The transition is efficient as long as the number of output fibres is equal to (or greater than) the number of modes supported in the multimode region. The first generation of lanterns were made by tapering down a bundle of single-mode fibres, all placed within the lower refractive-index preform, until their claddings and preform merged into a composite waveguide to become the core and cladding of a new multimode fibre \citep{Leon-Saval2005}. 
More recently, photonic lanterns have been demonstrated using a multi-core fibre (MCF) -- a single fibre containing many uncoupled single-mode cores, each effectively acting as its own single-mode fibre -- by placing it within a low refractive index capillary and tapering that down to form a single multimode core region \citep{Birks2012} (Figure \ref{fig_rsoftphasediag} (top)). This allows PLs with up to hundreds of output cores (and hence modes) to be manufactured \citep{Leon-Saval2017}, and the entire PL can fit entirely within a standard fibre connector. Crucially, the monolithic nature of the device where the mode conversion occurs (typically 20-60mm in length) means that, once manufactured, the relationship between the modes excited in the multimode region and the distribution of light in the uncoupled single-mode outputs is deterministic and unchanging. 

In the PL-WFS, the telescope PSF is injected directly into the multimode region of a photonic lantern. The photonic lantern then converts the multiple modes in the multimode fibre into an array of uncoupled single-mode outputs, with the distribution of flux between the outputs determined by the corresponding power in each mode at the input. Once in the form of single-mode cores, the information is robust -- it is encoded in only the intensity of each core, which is essentially unaffected by small perturbations. Moreover, when using a MCF, any wavelength-dependant loss and behaviour due to moderate bending and perturbation of the fibre will be the same across all cores.  
In the design presented here, the output of the lantern is in the form of a MCF. The distribution of power between modes can now be measured via single-pixel measurements of the flux in each waveguide, at a location remote from the focal plane. This enables the use of sensitive detectors (such as avalanche photodiodes) or wavelength dispersion onto an imaging detector to provide additional information.

In the end we have a stable system where we have $n$ intensity measurements (for an $n$ mode photonic lantern) which is a function of both the amplitude and phase of the telescope PSF. This transfer function can not be easily predetermined in manufacture due to fabrication imperfections, but it is fixed. If it can be learned, then it is possible to determine the phase and amplitude of the incident wavefront (to a degree of complexity determined by the number of modes measured). The learning of this transfer function and the subsequent prediction of the wavefront is made more difficult by the fact that (other than at very small wavefront errors) the relationship is non-linear, and so a conventional matrix-based approach is insufficient. Thus to perform this inference, a neural network is used, as described in Section \ref{sec_labsetup}.

\begin{figure*}
\centering
\includegraphics[width=0.9\textwidth,valign=c]{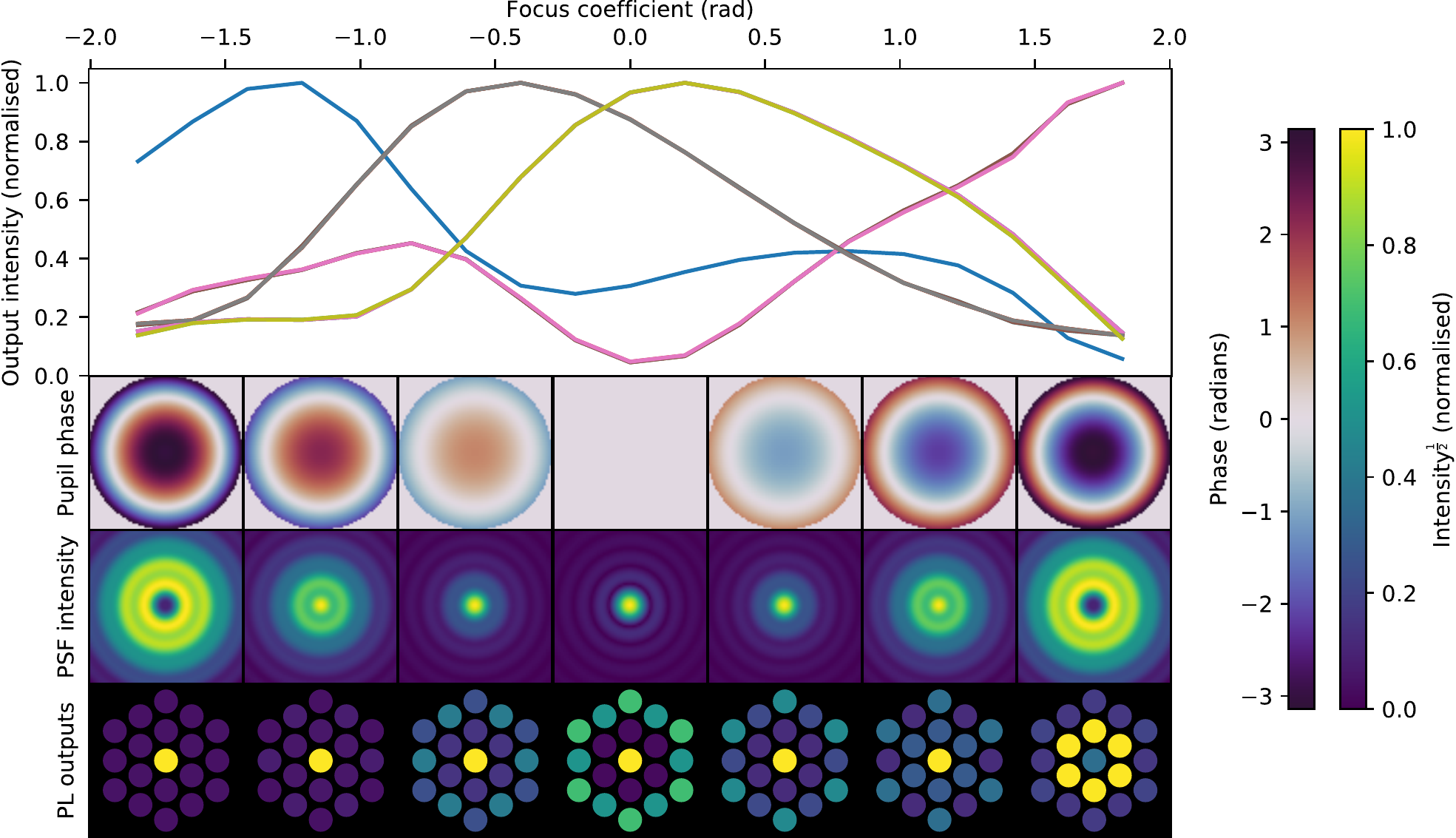}
    \caption{{\bf Photonic lantern wavefront sensor's non-degenerate response to a varying degree of defocus.
    } Results of simulations where a defocus term is applied and its amplitude scanned from -2 to +2 radians. In the top panel, the normalised output intensities of the the 19 single-mode outputs are plotted as a function of defocus amplitude (although only 4 separate trends are seen due to the symmetry of this aberration). In the lower three rows the pupil phase, PSF intensity and PL outputs are shown as per Figure \ref{fig_rsoftphasediag}. It is seen that although positive and negative defocus terms of the same amplitude give identical PSFs, it is unambiguous in the measurements from the PL. However, it is also seen that there is not a simple linear relationship between the amplitude of the phase error and the intensity of the lantern outputs. Simulations are performed at a wavelength of 1550~nm. Intensities are plotted with a square-root stretch to better show faint detail.
    }
    \label{fig_rsoftscandiag}
\end{figure*}

To validate the approach, a series of simulations were performed. First, a wavefront containing Zernike aberrations is produced and the complex electric field of the resulting PSF is 
obtained. This is then input into a model of the photonic lantern built using the {\ttfamily RSoft} software from Synopsis. Here, a numerical simulation is performed wherein the electric field is allowed to propagate from the multimode end to the single-mode outputs.

The result of one simulation demonstrating this concept is shown in Figure \ref{fig_rsoftphasediag} (bottom), wherein the phase of the wavefront, the intensity and phase of the resulting PSF after focusing, and the intensity of the 19 single-mode core outputs of the photonic lantern are given. The results for three wavefronts are shown -- one with a flat wavefront, and the other two with +0.8 radians and -0.8 radians of astigmatism respectively. It is important to note that, in the latter two cases, the intensity structure of the PSFs are identical, and so a conventional imaging sensor at the focal plane would not be able to distinguish them. However the necessary information is contained within the phase structure of the PSF, which is successfully measured by the photonic lantern and encoded in the intensity of its outputs. 

These numerical simulations also demonstrate the non-linear response of the lantern's output intensities to wavefront phase. In Figure \ref{fig_rsoftscandiag} a series of simulations are run where a defocus term of changing amplitude is applied, and the output intensities of the lantern plotted as a function of defocus amplitude. It is seen that the 19 output intensities are not a linear function of phase, suggesting that using a linear algorithm (such as used conventionally in adaptive optics) to reconstruct the input phase would perform poorly.

\subsection*{On sky application}
In one proposed on-sky application, the telescope PSF is focused onto the tip of the photonic lantern, and the emerging single-mode, multi-core fibre routed to a suitable detector location. In the most basic setup, the MCF output is re-imaged onto a sensitive high-speed array detector, such as an EMCCD or sCMOS camera. Optionally, a low-dispersion prism can be inserted to allow low-resolution spectral information to be obtained, potentially useful for more advanced wavefront control and telemetry algorithms, as well as science. The output of the MCF can be spectrally dispersed with no additional reformatting or slit, using the so-called photonic `TIGER' configuration \citep{Leon-Saval2012, Betters2014, Betters2020}. The output of this camera is then fed to the real-time computer of the adaptive optics system, where the incident wavefront error is inferred (using a simple neural network) and the appropriate correction applied to the deformable mirror.

Rather than acting as a stand-alone wavefront sensor, the same fibre can feed a high-dispersion single-mode spectrograph for science measurements. This offers some major advantages over a standard multimode-fibre-fed spectrograph, as described by various authors (e.g. \cite{Bland-Hawthorn2010, Betters2013, Harris2015, Jovanovic2016c, Crepp2016}), outweighing the cost (and potentially extra detector noise) arising from the extra pixels required. 
By converting the multimode light of the telescope PSF into a set of single-modes, the scaling relation between telescope aperture and the size of the spectrograph optics is broken.  
This vastly reduced size results in an instrument with far more stability (crucial for high-dispersion spectroscopy) and also allows multiple instances of the spectrograph to be easily replicated to allow a large number of objects to be simultaneously observed. Moreover, the spatial filtering intrinsic to a single-mode fibre removes the modal noise that limits the spectral stability of conventional spectrographs \citep{Baudrand2001}.
Photonic lantern enabled single-mode spectrographs are the focus of ongoing efforts and technology demonstrators (e.g. \cite{Feger2016,Gatkine2017,Jovanovic17}). Furthermore this technique has been successfully employed in other light-starved applications with high-stability requirements, such as Raman spectroscopy \citep{Betters2020}.

When the same lantern and fibre is used both as a WFS and to feed the spectrograph, a truly zero non-common-path design is realised. In this case, it is likely that a separate dispersing element and detector will be used for the science spectrograph than for the wavefront sensing portion. This is because very high dispersion spectrographs need very large detectors with very long integration times to reach the required signal/noise ratio,  
while the wavefront sensor needs to operate at a high framerate.
To enable this, a dichroic mirror can be placed within the re-imaging optics after the termination of the MCF, directing longer wavelength light to the appropriate dispersion and detection modules for science. However, a new generation of fast, low-noise infrared detectors using e-APD technology are now available \citep{Finger2014}, which may remove this requirement. One limitation of utilising a dichroic to split the light is the introduction of differential chromatic features between the WFS and science light (a problem not encountered in an imaging (rather than spectroscopic) FP-WFS application since the same imaging detector is used for all signals). If the effect proves to be large, then mitigation methods include trading photon-noise for differential chromaticity by using a grey beamsplitter instead, or performing PL-WFS measurements at both longer and shorter wavelengths than the science observation and interpolating the correction via a model.

Also, it is straightforward to build a multi-object wavefront sensor (e.g. for use in a multi-conjugate adaptive optics system \citep{Marchetti2007}) by simply adding more lantern/fibre units, and imaging the output cores from multiple MCFs onto a single larger detector or even multiple detectors.  
In the case of a multi-object galaxy survey, for example, the existing fiber positioning system could easily place multiple wavefront sensors where desired, since they have the same form factor as the existing multimode fibre infrastructure.

Another application is in the case of coronagraphic imaging. While the light from the region beyond the coronagraphic mask proceeds as usual to an imaging instrument, the (usually neglected) light reflected off the coronagraphic focal-plane mask or Lyot-stop could be redirected to the PL-WFS. The neural network architecture described here would be able to handle the distortion created by redirecting the light in this manner, since this is just a modification to the transfer function already being learned. This way multi-wavelength focal-plane wavefront sensing could still be performed while long-exposure science integrations take place.

In which of these configurations the PL-WFS is deployed depends on the science case. When the science object is not well spatially separated from the star, such as with radial-velocity measurements, transit spectroscopy, characterisation of circumstellar dust, etc., measuring the science data directly from the PL via high dispersion spectroscopy is ideal. For cases where the science object is well separated (such as a planet at several $\lambda/D$ separation), the planet would likely be outside the sensor's field of view, and the PL-WFS would be deployed purely as a focal plane WFS to optimise the performance of coronagraphic imaging or post-coronagraphic spectroscopy.

The number of modes supported in the multimode input of the PL is determined by its diameter. Since the PL-WFS is at the focal plane, its core diameter, and hence number of spatial modes, corresponds directly to its field of view. The number of single-mode outputs of the PL sets the limit on the number of spatial modes that can be sensed. The device demonstrated here uses a relatively small number of outputs (19) and hence number of modes, but this can be extended to higher order modes by increasing the number of outputs on the device. Currently, devices with up to 511 outputs \citep{Birks2015} are being produced. Since the outputs of the PL are orthogonal, the number of measurable spatial modes scales linearly with the number of outputs, however the optimal basis to be used for probing and/or reconstructing wave fronts with such a device is the topic of future work. 

Even a low mode-count device such as the current 19 output PL-WFS is extremely useful when used in the focal plane, since non-common-path-aberration is strongly dominated by low-order terms, with their amplitude very quickly diminishing as spatial frequency increases \citep{Sauvage2007}.  
Moreover, island modes / low wind effect modes are well represented by a low order mode set \citep{NDiaye2018}. Nonetheless, higher order non-common-path aberrations are also problematic (such as those arising from polishing error, sharp diffraction features, and other quasi-static aberrations), so the achievable Strehl ratio will be ultimately limited by the number of modes supported by the sensor.

\subsection*{Laboratory demonstration}
\label{sec_labsetup}

\begin{figure*}
\centering
\includegraphics[width=0.9\textwidth,valign=c]{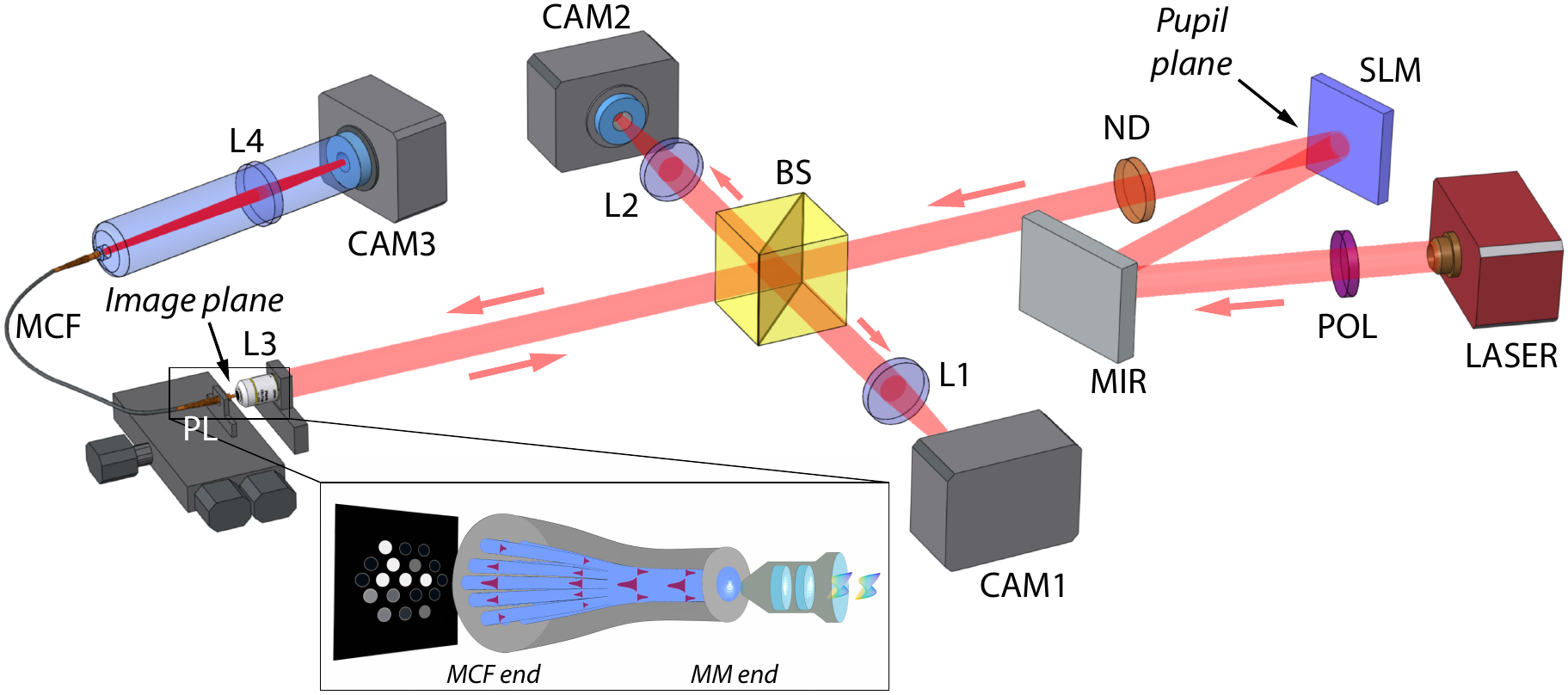} 
    \caption{{\bf Diagram of the laboratory setup used for testing the photonic lantern wavefront sensor.}  
    A collimated 685~nm laser (LASER) is passed through a linear polariser (POL) and via a fold mirror (MIR) onto a spatial light modulator (SLM), with a neutral density filter (ND) used to attenuate the beam. A wavefront constructed from a chosen set of Zernike terms is created by the SLM and focused to an image and injected by a microscope objective (L3) into the multimode end of the photonic lantern (PL). The intensity of the 19 outputs is then transmitted via multicore fibre (MCF) measured by a camera (CAM3) via lens L2. The raw PSF is also imaged via beamsplitter BS and lens L1 onto camera CAM1. The back-reflection off the fibre tip is imaged via the same beamsplitter and separate imaging system (L2, CAM2) to aid with alignment. Inset: illustration of the principle of the photonic lantern WFS. The incident aberrated wavefront is focused to an image at the focal plane, where the multimode end of the photonic lantern is placed. The complex wavefront determines the combination of of modes excited within the multimode region, which are then translated via an adiabatic taper transition into an array of single-mode outputs, the intensities of which encode the complex wavefront information.}
    \label{fig_explayout}
\end{figure*}

\begin{figure}
\includegraphics[width=0.48\textwidth]{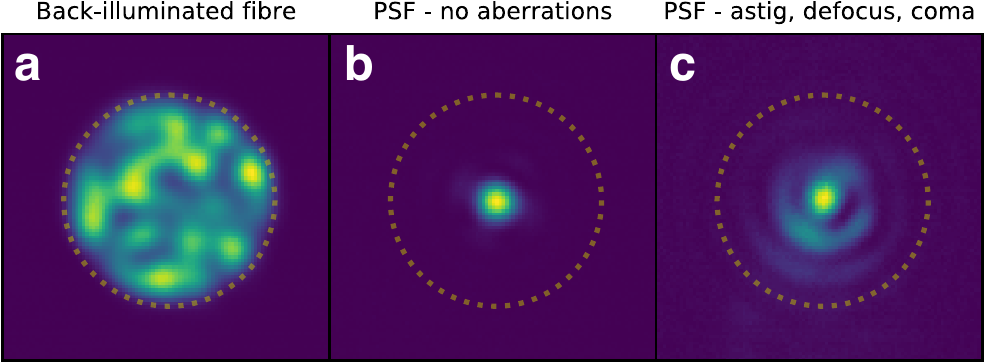} 
    \caption{{\bf Near-field image of the photonic lantern’s multimode end face.
    } The dotted line marks the outer extent of the fibre core, which has radius of approximately 3~$\lambda/D$. {\bf a} The lantern is back-illuminated by injecting light into the multi-core outputs (with random intensity distribution), exciting some combination of the fibre's modes, visible here. {\bf b} Back-reflected image of the multimode fibre when no aberrations are applied. {\bf c} Back-reflected image of the multimode fibre when several aberrations are applied.}
    \label{fig_backreflims}
\end{figure}

To validate the ability of the PL-WFS to determine the wavefront phase from the focal plane, a laboratory experiment was performed and the ability to recover the incident wavefront errors from the PL outputs was demonstrated. 
The experimental testbed provided the ability to inject a PSF arising from an arbitrary wavefront (created using a spatial-light modulator (SLM)) into a photonic lantern, and measure the 19 output intensities. The experimental layout is shown in Figure \ref{fig_explayout}; see `Methods' for a detailed description.  
A set of images produced by the back-reflection imaging system, showing the input face of the lantern and the back-reflected PSFs, are shown in Figure \ref{fig_backreflims}.

As seen in Section \ref {sec_theory}, 
the relationship between the input wavefront phase and the output intensities is not linear (or even monotonic for large phase errors). This means that reconstructing the input wavefront from the output intensities using a linear algorithm, such as the SVD-based approach conventionally used in adaptive optics, is not optimal. 
To address this, a multi-layer neural network was implemented, and various architectures tested. It was then trained and validated using laboratory data produced using the aforementioned laboratory setup. As a point of comparison, a traditional linear, singular-value-decomposition (SVD) based approach was also tested. See `Methods' for further details.

For each laboratory measurement, a combination of the first 9 (non-piston) Zernike terms are simultaneously applied to the SLM, each with an amplitude randomly chosen between approximately -0.12$\pi$ and 0.12$\pi$ radians. 
After these aberrations are combined the resulting phase error for each measurement has a peak-to-valley amplitude of approximately $\pi$ radians. 
This is a limit imposed by the maximum retardance the SLM can produce within its linear range. 

The 19 output intensities from the photonic lantern are then recorded, and the images of the PSF and back-reflection from the fibre are also saved for reference. This is then repeated for the desired number of samples. For the results in this paper, a data set of approximately 60000 measurements was taken, which would take of order 30 seconds to acquire with a contemporary extreme AO system running at kHz speeds.  
Of these data, 20\% are reserved as validation samples and the rest are used as training samples. To evaluate the performance of the network, the 19 output fluxes for previously unseen laboratory test data were given to the neural network and the wavefront coefficients predicted, and the mean-squared error between the predicted coefficients and the true coefficients calculated.

The neural network was able to reconstruct the incident wavefront error to varying degrees of accuracy depending on the model architecture chosen; a few representative models and their root-mean-squared-errors are given in Table \ref{table_lossfuns}. It was clear that a non-linear architecture is needed. The best performing network (using the non-linear, ReLU activation function) yielded a root mean squared error (RMSE) of just $5.1\times10^{-3}\;\pi$ radians, 
while the traditional linear approach (using the singular value decomposition method) gave a much worse RMSE of $3.0\times10^{-2}\;\pi$ radians. 

It was also found that a deep network (i.e. including hidden layers) was required for optimum performance. The best performing network mentioned above (RMSE = $5.1\times10^{-3}\pi$) 
consisted of 3 layers arranged in a `funnel' configuration, with each layer having 2000, 1050 and 100 units respectively. A single layered network (with 2000 units) shows worse performance, with an RMSE of $7.6\times10^{-3}\pi$.  
Furthermore, it was found that while performance was sensitive to the number of units in the first layer(s) and the number of layers, it was quite insensitive to the number of units in the final layer(s); increasing the number of units in the final layer beyond 100 had little effect. Increasing the number of hidden layers beyond 3, or the number of units in the first layer beyond 2000, also gave rapidly diminishing returns. Regularisation using dropout was also tested, but had little effect except for with very large networks (>3000 units in the first layer, or >3 layers), but which still offered no improvement over the smaller networks described above. These values are produced from a model trained on the complete set of data (48 000 individual measurements). But useful results are found even with much less data; training with 4 800 measurements gives an RMSE of $9.3\times10^{-3}\pi$  
and with only 480 measurements gives an RMSE of $2.0\times10^{-2}\;\pi$ radians.

\begin{table} 
\footnotesize
\begin{tabular}{lllll}
\hline
\multicolumn{1}{c}{Activation} & \multicolumn{1}{c}{Neurons in}  & \multicolumn{1}{c}{Neurons in}  & \multicolumn{1}{c}{No. of}     & \multicolumn{1}{c}{RMS error}       \\
                               & \multicolumn{1}{c}{first layer} & \multicolumn{1}{c}{final layer} & \multicolumn{1}{c}{hidden layers} & \multicolumn{1}{c}{$\times10^{-3}\;\pi$ radians} \\ \hline
Non-linear                     & 2000                            & 100                             & 2                                  & \textbf{5.1}                           \\
(ReLU)                         & 2000                            & 2000                            & 2                                  & \textbf{5.1}                           \\
                               & 2000                            & 100                             & 1                                  & \textbf{5.9}                           \\
                               & 200                             & 30                              & 6                                  & \textbf{6.4}                           \\                               
                               & 2000                            & -                               & 0                                  & \textbf{7.6}                           \\
                               & 100                             & 100                             & 3                                  & \textbf{7.6}                           \\
                               & 100                             & -                               & 0                                  & \textbf{17}                            \\ \hline
Linear                         & -                               & -                               & -                                  & \textbf{30}                           
\end{tabular}
\caption{The performance of several different neural network architectures (selected from a larger hyperparameter scan) in predicting the incident wavefront error from the 19 PL output fluxes, quantified by the root-mean-squared-error (in $\pi$ radians) of the predictions using test data. A deep, funnel-shaped network gives the lowest error. The ability of a neural network to handle non-linearity is clearly advantageous. See text for details.}
\label{table_lossfuns}
\end{table}

\begin{figure} 
	\includegraphics[width=0.47\textwidth,valign=c]{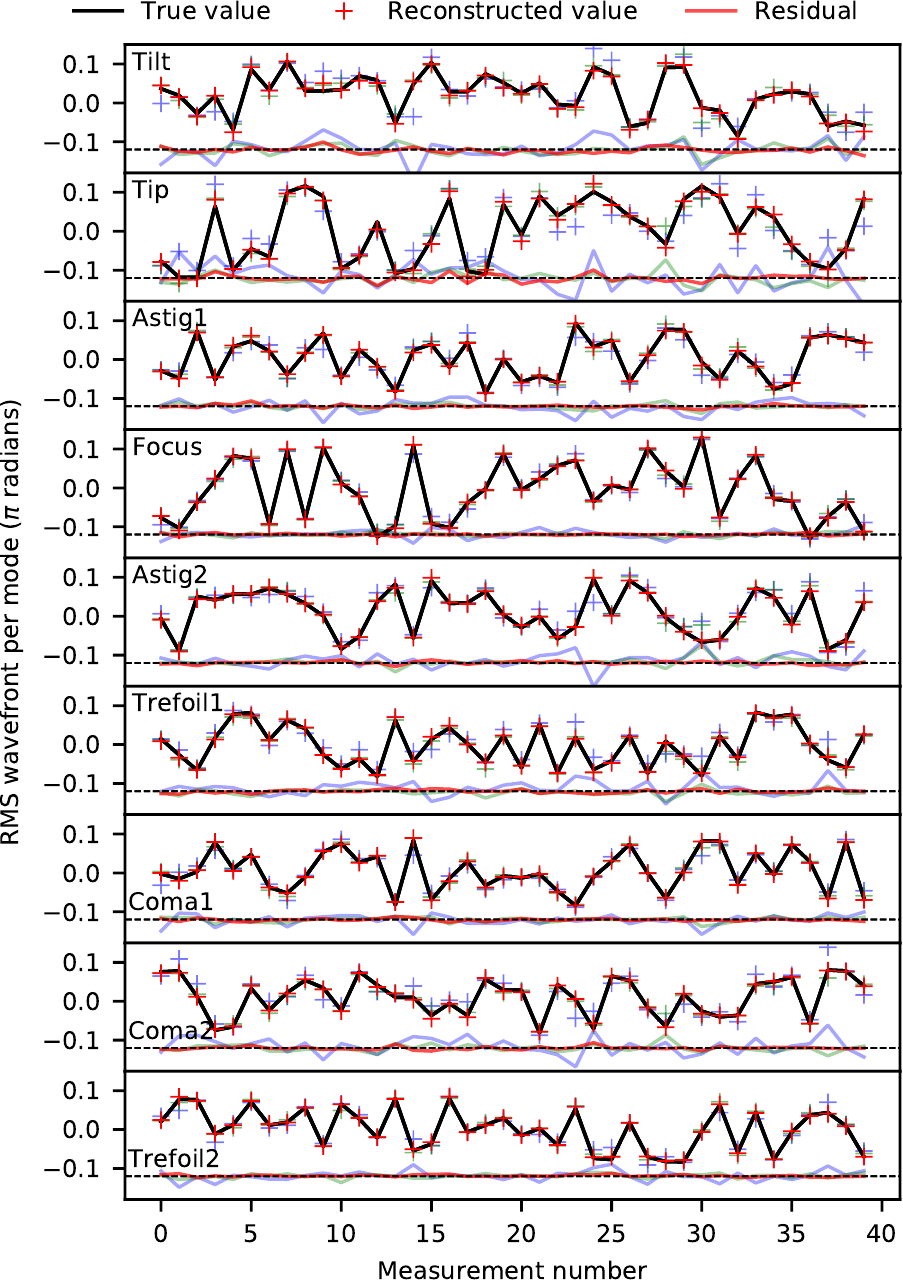}
    \caption{{\bf Results of laboratory tests of the photonic lantern wavefront sensor.} Shown here are the predicted Zernike coefficients (crosses) and the true values (black lines) for a randomly selected set of 40 measurements. Red points are predictions from a model trained with 48000 measurements, green points with 4800 measurements and blue with 480 measurements. The difference between the predicted and true values is plotted at the bottom of each panel. Each measurement consists of a combination of the first 10 Zernike terms each with a randomly chosen amplitude between approximately -0.12$\pi$ and 0.12$\pi$ radians applied to the SLM. Resulting combined wavefronts for each measurement have peak-to-valley amplitudes of order $\pi$ radians (limited by SLM hardware). Predictions are performed by the neural network described in the text, using the 19 output intensities of the lantern. The neural network accurately predicts the Zernike terms of the wavefront injected into the lantern, with a root-mean-squared-error of $5.1\times10^{-3}\;\pi$ radians.
    } 
    \label{fig_fittedmodel}
\end{figure}

Figure \ref{fig_fittedmodel} shows the results of the wavefront reconstruction (in terms of the 9 labelled Zernike modes) for laboratory data using the best model architecture. Data for 40 randomly selected samples are shown, with the reconstructed wavefront coefficients overplotted on the true values. It is seen that for all terms the reconstructed values align extremely well with the true values, with little deviation. Interestingly the tip and tilt terms show the poorest performance. This is believed to be due to drift in the alignment of the laboratory setup (due to thermal drift) as training data was acquired, leading to positional modes being poorly learned. 

This experiment was performed with a narrow-band light source (bandwidth 1.2~nm), while in astronomy a much broader bandwidth would be desired for increased photon efficiency. As described previously, the anticipated implementation would be spectrally dispersed, either at high spectral resolution for simultaneous science spectroscopy or at low resolution for wavefront-sensing only.
As an individual spectral channel becomes broader (beyond that seen in this experiment), the light's coherence, and hence the degree of modulation of the PL outputs, decreases. This would be expected to lead to a gradual limitation in sensitivity, and the optimal balance between channel width, read noise (from increased spectral dispersion) and total bandwidth is the subject of future analysis.

\section*{Discussion}
\label{sec_conclusion}
The photonic lantern wavefront sensor (PL-WFS) represents a type of wavefront sensor which addresses several of the limitations of current adaptive optics systems. Placing the wavefront sensor at the focal plane, rather than at a non-common pupil plane, has been long desired in adaptive optics as it eliminates non-common path error and is sensitive to wavefront errors not visible in the pupil plane (such as island modes). However the image at the focal plane does not contain sufficient information for wavefront reconstruction, since it contains only intensity information and lacks the phase component, leading to degeneracies. Other focal-plane wavefront sensor designs rely on introducing further perturbations to the wavefront to break degeneracies, linear approximations (so unsuited to large phase error) or slow, non-real time methods. They also are poorly suited to injecting the image into single mode fibers, extremely important for major science goals such as spectrographic characterisation of exoplanet atmosphere.

The PL-WFS addressees these limitations by placing the multimode region of a photonic lantern at the focal plane, which deterministically remaps the combination of mode-fields in the multimode region to a set of intensities produced at several single-mode outputs. Since the modes excited in the multimode region are a function of both the amplitude and the phase of the incident wavefront, non-degenerate wavefront information is contained and the wavefront can be reconstructed. Furthermore, since the light is optimally injected into single-mode fibres, it is ideal for subsequent connection to a single-mode spectrograph. To deal with the non-linear relationship between phase and intensity in this device, a neural network is employed. 

Simulations validate the principle of the device, and laboratory demonstrations confirm its operation. In laboratory tests, wavefront errors with P-V amplitude of $\sim\pi$~radians 
constructed using the first 9 (non-piston) Zernike terms are introduced, and are then accurately reconstructed from a focal plane measurement using the PL-WFS, to a precision of $5.1\times10^{-3}\;\pi$ radians root-mean-squared-error. 

The next steps are to use the device in a closed-loop configuration wherein wavefront errors are corrected in real-time, and introduce wavefront errors using a basis more similar to that of a turbulent media (such as a Kolmogorov phase screen). Following that, the device can be tested in an on-sky deployment at an astronomical telescope. Eventually the PL-WFS will  
form a key component in the increasingly complex set of sensors within a modern adaptive optics system, paving the way for advanced imaging and characterisation of exoplanets, their atmospheres and surface composition, and the detection of biological signatures.

\section*{Methods}
\subsection*{Laboratory procedure}
The experimental layout is shown in Figure \ref{fig_explayout}, which allows the PSF produced by an arbitrary wavefront to be injected into the photonic lantern, and its output fluxes measured. Also, to aid in the alignment of the lantern, a back-reflection imaging system was implemented wherein the end of the multimode region is directly imaged via the same lens as that used for injection, with the incident PSF visible via its reflection off the polished end of the fibre. An example of these images is given in Figure \ref{fig_backreflims}. 
For these images, light was simultaneously injected into the single-mode outputs of the lantern, to excite a combination of modes in the multimode region. The superposition of these modes are seen in the left panel of the figure as the speckle-like background pattern in the fibre core. 
A separate focal plane camera was also implemented to independently verify the PSF of the system for a given SLM-induced wavefront.

A 685~nm laser with measured bandwidth 1.2~nm 
is injected into a single-mode fibre and collimated by an off-axis parabolic mirror, followed by a 4~mm diameter pupil stop. The beam passes through a linear polariser (the SLM operates in a single linear polarisation) and onto the SLM via a fold mirror. From the SLM it passes through a neutral density filter to the beam-splitter cube (non-polarising, R:T 50:50). Here the reflected 50\% of the beam is focused onto to the imaging camera (FLIR Grasshopper3 - CAM1) via an f=200~mm doublet lens to provide a PSF reference, while the transmitted beam is focused onto the tip of the multimode region of the photonic lantern via a 10x microscope objective lens.

The PL used here is made using a visible wavelength multi-core fibre (MCF) with 19 uncoupled cores with a 3.7~\textmu m core diameter, NA of 0.14, and core-to-core separation of 35~\textmu, instead of a bundle of SMFs \citep{Leon-Saval2013, Birks2015}, that is tapered with a low-index glass capillary (fluorine doped fused silica) jacket to produce a 22~\textmu m MM input with an NA of 0.145. This is equivalent to an angular `field-of-view' with radius of approximately 3~$\lambda/D$. 
The PL is then housed within a standard SMA fibre connector.
The lantern is mounted on a 3-axis stage to align it with the PSF. The output of the multicore fibre is then imaged onto a separate camera (FLIR Chameleon3 - CAM3) via an f=200~mm doublet lens, to record the flux in each of the 19 single-mode outputs. Meanwhile, the back-reflected light from the multimode fibre tip (arising from the Fresnel reflection of the non-AR coated fibre) passes back through the microscope objective and is focused onto another camera (FLIR Blackfly - CAM2) via a reverse pass through the same beamsplitter cube, to aid with alignment.

Each measurement of the multicore outputs was performed with 10 co-adds of 20~ms integrations, with this relatively long total integration time required to smooth out the ripple caused by the SLM's refresh rate. To limit the effect of drifting alignment, the experimental setup was placed in a temperature-stabilised room, maintaining the temperature to within $\pm 0.1^\circ$ C. All wavefront modulation and data capture is performed via Matlab.

\subsection*{Data analysis}
The neural network was implemented using Keras \citep{Chollet2015}, using the Tensorflow backend \citep{tensorflow2015-whitepaper}. The loss function used was the mean squared error of the predicted coefficients, and using a ReLU activation function and Adam optimizer.  A range of architectures for the neural network was explored, with hyperparameter exploration and optimisation performed using Talos \citep{Talos2019}. 

In addition to the neural network, a linear, singular-value-decomposition (SVD) based approach (traditionally used in adaptive optics) \citep{Guyon2018} was tested as a point of comparison. Here, a matrix is constructed mapping the input wavefront coefficients to the output intensities using the training data, and then a pseudo-inverse of the matrix is create using a SVD, with suitable regularisation. This pseudo-inverse matrix is then used to predict the wavefront coefficients from any set of previously unseen output fluxes.




\section*{Acknowledgements}
We would like to thank Prof Birks and Dr Gris-Sanchez from the University of Bath for facilitating the fibre fabrication and the use of the fibre drawing tower. S.G.L-S would like to thank A/Prof Amezcua-Correa from the College of Optics and Photonics (CREOL) at the University of Central Florida for the inspiring conversations about this research and possible applications outside astronomy.

\section*{Author contributions}
B.R.M.N. and S.L-S. developed and led the project.  C.H.B. designed and fabricated the multicore photonic lantern and S.L-S. designed and fabricated the multicore fibre.  J.W. conducted the laboratory experiments with assistance from B.R.M.N., C.H.B. and S.L-S.  B.R.M.N. developed and applied the deep learning network and algorithms.  B.R.M.N., A.W. and J.W. analysed the data.  B.R.M.N. wrote the paper with contributions from all the authors. S.L-S. and B.R.M.N. supervised the study.

\section*{Competing interests}
The authors declare no competing interests.

\section*{Corresponding author}
Correspondence to Barnaby Norris

\section*{Data availability statement}
The data produced in this study are available from the corresponding author upon reasonable request.


\label{lastpage}
\end{document}